\documentclass{llncs}

\usepackage[utf8]{inputenc}
\usepackage{amsmath}
\usepackage{amsfonts}
\usepackage{amssymb}
\usepackage{booktabs} 
\usepackage{textcomp}
\usepackage{hyperref}
\usepackage{pgf}
\usepackage{tikz}
\usetikzlibrary{shapes.multipart}
\usetikzlibrary{arrows,positioning,automata}
\usetikzlibrary{shapes.geometric}
\usepackage{listings}
\usepackage{xcolor}
\usepackage{color}
\usepackage{bussproofs}
\usepackage{stmaryrd}
\usepackage{graphicx}
\usepackage[normalem]{ulem}

\usepackage[final,layout=footnote,mode=multiuser]{fixme}
\fxusetheme{color}
\FXRegisterAuthor{gc}{angc}{GC}
\FXRegisterAuthor{lg}{anlg}{LG}
\FXRegisterAuthor{ac}{anac}{AC}

\lstdefinestyle{CPSL}{
  breaklines=true,
  xleftmargin=\parindent,
  language=python,
  showstringspaces=false,
  basicstyle=\small\ttfamily,
  keywordstyle=\bfseries\color{blue!40!black},
  commentstyle=\itshape\color{gray!40!black},
  stringstyle=\color{purple},
  escapeinside={(*}{*)},
  morekeywords={STATE, GIVEN, WHEN, THEN, ACCEPT}
}

\definecolor{verylightgray}{rgb}{.97,.97,.97}

\lstdefinelanguage{Solidity}{
	keywords=[1]{anonymous, assembly, assert, balance, break, call, callcode, case, catch, class, constant, continue, constructor, contract, debugger, default, delegatecall, delete, do, else, emit, event, experimental, export, external, false, finally, for, function, gas, if, implements, import, in, indexed, instanceof, interface, internal, is, length, library, log0, log1, log2, log3, log4, memory, modifier, new, payable, pragma, private, protected, public, pure, push, require, return, returns, revert, selfdestruct, send, solidity, storage, struct, suicide, super, switch, then, this, throw, transfer, true, try, typeof, using, value, view, while, with, addmod, ecrecover, keccak256, mulmod, ripemd160, sha256, sha3}, 
	keywordstyle=[1]\color{blue}\bfseries,
	keywords=[2]{address, bool, byte, bytes, bytes1, bytes2, bytes3, bytes4, bytes5, bytes6, bytes7, bytes8, bytes9, bytes10, bytes11, bytes12, bytes13, bytes14, bytes15, bytes16, bytes17, bytes18, bytes19, bytes20, bytes21, bytes22, bytes23, bytes24, bytes25, bytes26, bytes27, bytes28, bytes29, bytes30, bytes31, bytes32, enum, int, int8, int16, int24, int32, int40, int48, int56, int64, int72, int80, int88, int96, int104, int112, int120, int128, int136, int144, int152, int160, int168, int176, int184, int192, int200, int208, int216, int224, int232, int240, int248, int256, mapping, string, uint, uint8, uint16, uint24, uint32, uint40, uint48, uint56, uint64, uint72, uint80, uint88, uint96, uint104, uint112, uint120, uint128, uint136, uint144, uint152, uint160, uint168, uint176, uint184, uint192, uint200, uint208, uint216, uint224, uint232, uint240, uint248, uint256, var, void, ether, finney, szabo, wei, days, hours, minutes, seconds, weeks, years},	
	keywordstyle=[2]\color{teal}\bfseries,
	keywords=[3]{block, blockhash, coinbase, difficulty, gaslimit, number, timestamp, msg, data, gas, sender, sig, value, now, tx, gasprice, origin},	
	keywordstyle=[3]\color{violet}\bfseries,
	identifierstyle=\color{black},
	sensitive=false,
	comment=[l]{//},
	morecomment=[s]{/*}{*/},
	commentstyle=\color{gray}\ttfamily,
	stringstyle=\color{red}\ttfamily,
	morestring=[b]',
	morestring=[b]"
}

\newcommand{\projname}{PrYVeCT}

\begin{document}
\title{Private-Yet-Verifiable Contact Tracing\\{\normalsize Short paper}}

\author{Andrea Canidio \and Gabriele Costa \and Letterio Galletta}
\institute{IMT School for Advances Studies}

\maketitle

\begin{abstract}
We propose \projname{}, a \textit{private-yet-verifiable} contact tracing system. 
\projname{} works also as an authorization framework allowing for the definition of fine-grained policies, which 
a certain facility can define and apply to better model its own access rules.
Users are authorized to access the facility only when they exhibit a contact trace that complies with the policy. 
The policy evaluation process is carried out without disclosing the personal data of the user. 
At the same time, each user can prove to a third party (e.g., a public authority) that she received a certain authorization. 
\projname{} takes advantage of \emph{oblivious} automata evaluation to implement a privacy-preserving policy enforcement mechanism.
\end{abstract}


\section{Introduction}
\label{sec:introduction}

Many countries are responding to the current COVID-19 epidemics by collecting their citizens' location data, via smartphone apps. 
The purpose is to automate the process of contact tracing: when a person is found positive to the virus, all people who came in close contact with her during the previous days are asked to enter self quarantine. 
The success of some Asian countries in containing the spread of the virus is often ascribed to this massive data collection effort. 
Despite the success, this data collection raises serious privacy concerns~\cite{harari2020world}.

As a response, researchers have started working on \emph{fully-private} contact-tracing apps~\cite{tang2020privacy}.
Roughly speaking, these proposals collect anonymously location and contact data  of users via Bluetooth and store them locally.
When a user is found positive to the virus, the contact data of the infected person are uploaded on a central server. 
Each app periodically checks the data uploaded on the server against the data  collected in the previous two weeks. 
If a match is found, the user is notified of a dangerous encounter, and she is invited to  enter self quarantine for a certain amount of time. 
Since the focus of these proposal is on total privacy, it is impossible for the public authority to determine whether a given person is indeed complying with the mandatory quarantine.
This is an important limitation because even a single infected person who does not comply may be enough to spread the virus~\cite{nytimes}. 
Hence,  non-compliance may make fully-private contact tracing ineffective at controlling the pandemic.


In this paper, we propose \projname{} a contact tracing system that is private yet verifiable. 
Private because as previous proposals, e.g.,~\cite{troncoso2020decentralized}, it does not exposes any of the collected personal data.
Verifiable because each user can prove to have received an authorization to circulate or to enter a particular facility.
%
%



The main novelty of our approach resides in an authorization infrastructure built on user the past history of user (her \emph{trace}) which includes her contacts, 
location data, and other relevant events, e.g., recent swab test results. 
Based on this history, the authorization infrastructure provides each user with a token whose validity can be verified, e.g., by the public authority.
Thus, using this mechanism a \emph{distributed history-based access control} (DHBAC) framework~\cite{BartolettDF05} can be implemented, ensuring that whenever an authorization token is issued to an individual, her history complies with a given  authorization policy. 


The rest of the paper is organized as follows.
Section~\ref{sec:overview-example} introduces a motivating example for our proposals. 
An overview of \projname{} is in Section~\ref{sec:overview}. 
Section~\ref{sec:policy} introduces our policy specification language.
Section~\ref{sec:details} describes how the authorization and the enforcement mechanisms of \projname{} work.
Section~\ref{sec:literature} compares our proposal with the literature.
Section~\ref{sec:issues} concludes discussing some open issues and possible extensions.

\section{Motivating Example}
\label{sec:overview-example}

Several access control policies have been proposed by the public authorities in order to counter the spreading of the COVID-19 epidemic.
For instance, quarantine means forbidding the access to the outside, public spaces for a certain period of time.
Typically, the authorization to access a certain area/facility is granted/revoked according to certain, well defined events such as the outcome of an antibody test.
As an example, we propose the following two policies expressed in natural language.

\begin{enumerate}
\item  ``Citizens who could be contagious should not leave their residence until they are negative to a swab test''
\item ``Students that had $n$ or more risky contacts (i.e., contacts with someone positive to COVID-19) should not be admitted at school until they are negative to a viral test''
\end{enumerate}

Both of them regulate the access to a facility, i.e., the outside and a school (respectively).
Also, in both of them, the access authorization is granted/revoked as a consequence of certain events occurred in the past.

Assuming that a certain authority is in charge for defining and applying a certain policy, some considerations are necessary.
In terms of privacy, the enforcement of policies such as the previous ones may require to reveal (part of) the past behavior of a citizen, e.g., the list of personal contacts or the result of a clinical test.
Moreover, in terms of effectiveness, once an authorization is issued, it should be verifiable by all the interested parties.
The reason is that, in some cases, a distributed monitoring mechanism can be implemented through the enrollment of third parties.
For instance, the authorization associated to the first policy above can be verified by both a police officer and a private citizen.

\section{System Overview}
\label{sec:overview}

\begin{figure}[t]
\centering
\includegraphics[width=0.9\textwidth]{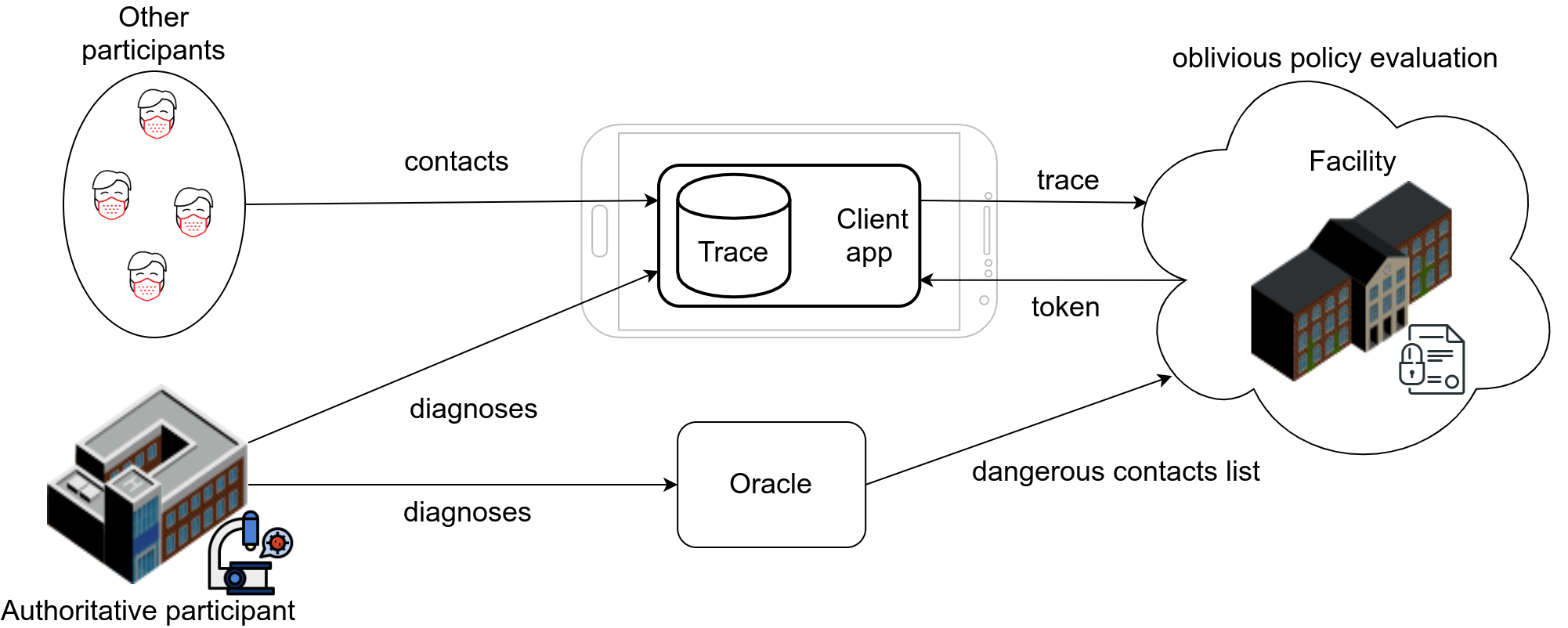}
\caption{Overview of the approach.}
\label{fig:overview}
\end{figure}

In this section we present an overview of our approach.
Figure~\ref{fig:overview} schematically depicts it.
The main stakeholders are the following.

\noindent
\emph{Users}. They participate to the contact tracing and policy enforcement system. 
Users are enrolled when they install a certain mobile app on their smartphone.
Each user is assigned to an anonymous identifier, i.e., a code that cannot be used to infer the actual identity of its owner.
The mobile app is responsible for recording the trace of events that have to do with the user.
These events include contacts with other users, e.g., generated when another device enters the Bluetooth transmission range, as well as other events triggered by an authoritative participant (see below).
Finally, the mobile app is responsible for interacting with the privacy-preserving policy evaluation protocol.

\noindent
\emph{Authoritative participants}. These participants represent entities having a specific role in the system.
For instance, a clinic that performs tests.
These participants generate special events, such as the outcome of a swab test.
These events are transmitted to the involved users, e.g., the citizen the test refer to, and also stored anonymously in a oracle server.

\noindent
\emph{Facilities}. A facility is any space that requires an access authorization. 
For instance, facilities may include school, train stations and even the outside (as in the quarantine example).
Facilities are assigned to a policy.
When a user request to access the facility, the authorization is granted if and only if the user's trace complies with the policy.
Eventually, if authorized, a user is provided with a verifiable token.

\noindent
\emph{Oracle server}. The oracle is a remote database where anonymous data is stored.
Authoritative participants store anonymous data that are needed to perform a policy evaluation.
The oracle server is queried by facilities during the policy evaluation process.
In particular, the policy evaluation process relies on the oracle for checking whether a trace contains dangerous contacts.
This is done without disclosing the identity of the involved parties.

\section{Policy framework}
\label{sec:policy}



\paragraph{Trace encoding.}


A trace $\sigma$ is a finite sequence of \emph{event} symbols taken from a finite alphabet $\Lambda$.
We use small letters $a,b$ to denote an event.
Moreover, given a finite set of labels $L$ and finite set of values $V$, we write $a(v)$ (for any $a \in L$ and $v \in V$) as a shorthand for the symbol $a_v \in L \times V$.
We write $\sigma \cdot \sigma'$ for the concatenation of two traces.
Also, if clear from the context, we feel free to omit $\cdot$, e.g., when writing $abc$ instead of $a\cdot b\cdot c$.

\begin{example}
\label{ex:trace}
Consider $L_t = \{a, s, v\}$ and $V_t = \{+,-\}$.
We use the alphabet $\Lambda_t = L \times V$ to denote the events corresponding to the outcome (i.e., positive $+$ or negative $-$) of \emph{antibody} ($a$), \emph{swab} ($s$) and \emph{viral} ($v$) tests.
For instance, $a(-)$ is the event corresponding to a negative antibody test result.
\end{example}

%
%
%
%


%
%
%

\paragraph{Policy specification language.}

Here we introduce our \emph{contact policy specification language} (CPSL).
CPSL inspires to both ConSpec~\cite{Aktug08conspec} and Gherkin~\cite{Wynne12cucumber}.
The syntax of CPSL policies follows the grammar of below.
\vspace{-14pt}
\begin{small}
\begin{center}
\begin{tabular}{l l}
\begin{tabular}{r @{ ::= } l}
$Pol$ & $State$ $Rules$ $Auth$\\
$State$ & \verb|STATE| $Decl$ \\
$Decl$ & $\lambda$ $\,\vert\,$ $Var$ : $Type$ \verb|:=| $Val$ $Decl$ \\
$Var$ & (variable names) \\
$Type$ & \verb!Id! $\,\vert\,$ \verb!Int! $\,\vert\,$ \verb!Bool! $\,\vert\,$ $Set$ \\
$Val$ & (constant values) \\
$Set$ & \verb!Set of! (\verb!Id! $\,\vert\,$ \verb!Int! $\,\vert\,$ \verb!Bool!)
\end{tabular} &
\begin{tabular}{r @{ ::= } l}
$Rules$ & $\lambda$ $\,\vert\,$ $Rule$ $Rules$ \\
$Rule$ & \verb|GIVEN| $Exp$ \verb|WHEN| $Event$ \verb|THEN| $Updt$ \\
$Exp$ & (boolean expressions) \\
$Event$ & (contact trace events) \\
$Updt$ & (assignments to state variables) \\
$Auth$ & \verb|ACCEPT| $Exp$
\end{tabular}
\end{tabular}
\end{center}
\end{small}
\vspace{-8pt}

Briefly, a policy ($Pol$) is a state declaration ($State$) followed by a finite sequence of rules ($Rules$) and an acceptance condition ($Auth$).
The state is a (possibly empty, denoted by $\lambda$) list of typed variable declarations ($Decl$).
Supported types ($Type$) are identifier (\verb!Id!), integer (\verb|Int|), boolean (\verb|Bool|) and sets ($Set$).
Each rule ($Rule$) is a triple of the form given-when-then.
The first element is a boolean expression defining the conditions under witch the rule applies (for brevity we omit writing \verb|GIVEN true|).
The ``when'' element contains the event that activates the rule.
If the same rule applies to several events, we list them in the ``when'' expression.
Finally, ``then'' contains a list of updates ($Updt$), i.e., assignments that change the state variables (where \verb|skip| stands for no modification).
The acceptance condition ($Auth$) is a boolean expression that characterizes the accepting states of the policy.
To clarify, we propose the following example.

\begin{example}
\label{ex:policy}

\begin{figure}[t]
\begin{tabular}{l @{\hspace{5pt}} l}
\begin{lstlisting}[morekeywords={Set, of, Id, Bool, Int, skip},xleftmargin=.2\textwidth, xrightmargin=.2\textwidth]
STATE
 contagious : Bool := false
  
GIVEN not contagious
WHEN a(+), s(+), v(+)
THEN contagious := true

GIVEN not contagious
WHEN a(-), s(-), v(-)
THEN skip
\end{lstlisting}
&
\begin{lstlisting}[morekeywords={Set, of, Id, Bool, Int, skip},xleftmargin=.2\textwidth, xrightmargin=.2\textwidth]
GIVEN contagious
WHEN s(-)
THEN contagious := false

GIVEN contagious
WHEN a(-), v(-), a(+), s(+), v(+)
THEN skip

ACCEPT not contagious
\end{lstlisting}
\end{tabular}
\caption{CSPL for the first policy of Section~\ref{sec:overview-example}.}
\label{fig:cspl}
\end{figure}

Figure~\ref{fig:cspl} shows the CPSL encoding of policy 1 of Section~\ref{sec:overview-example}.
The policy has one boolean variable, i.e., \verb|contagious|, and it consists of four rules.
The first rule states that \verb|contagious| is set to \verb|true| whenever a positive test event occurs.
The second rule states that negative tests do not change the current state when \verb|contagious| is false.
On the other hand, if \verb|contagious| is \verb|true|, it is set to \verb|false| when a negative swab test \verb|s(-)| occurs (third rule).
Instead, for any other event, the value of \verb|contagious| is not changed (fourth rule).
Finally, the acceptance condition is that \verb|contagious| is equal to \verb|false|.
\end{example}

\paragraph{Policy compliance.}

The semantics of a CPSL policy is given in terms of deterministic finite state automata (DFA).
A DFA consists of a tuple $A = \langle Q, q_0, \Lambda, \delta, F \rangle$ where $Q$ is a finite set of states (being $q_0$ the initial one), $\Lambda$ is an alphabet of symbols, $\delta : Q \times \Lambda \rightarrow Q$ is a transition function.
A computation amounts to iteratively apply the transition function to the elements of a trace $\sigma \in \Lambda^\ast$.
At each step the automaton is in a state $q$, being initially $q_0$. 
Then, the next symbol $a$ of $\sigma$ is evaluated to obtain the next state $q' = \delta(q,a)$.
A trace $\sigma$ is accepted by the automaton $A$ if the last state reached after the evaluation of all the symbols of $\sigma$ belongs to $F \subseteq Q$, i.e., the set of final states.

The translation from CPSL to the corresponding DFA is analogous to that presented in~\cite{Aktug08conspec}.
Intuitively, the automaton has a state for each valid assignment of values to the CPSL variables, where $q_0$ is for the initial assignment.
Transitions are created for each policy rule.
In particular, a transition is added for each state and symbol that satisfies the ``given'' expression.
The target state is obtained by applying the ``then'' updates to the variables of the source state.
Finally, all the states the satisfy the CPSL \verb|ACCEPT| expression are added to $F$.
To clarify, we propose the following example.

\begin{example}
\label{ex:semantics}
Consider again the same policy give in Example \ref{ex:policy}.
Intuitively, the corresponding DFA only has two states, i.e., $q_0$ (for \verb|contagious| equal to \verb|false|) and $q_1$ (for \verb|contagious| equal to \verb|true|).
Also, it is easy to verify that $F = \{q_0\}$.

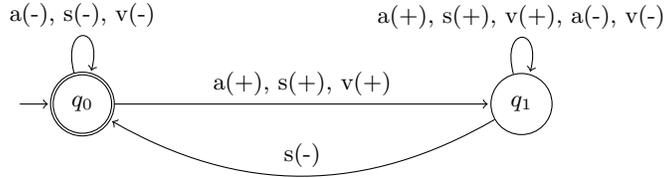
\begin{figure}[t]
\centering
\begin{tikzpicture}[shorten >=1pt,node distance=5cm,initial text={}]
  \node[state,initial,accepting]   (q_0)                {$q_0$};
  \node[state] (q_1) [right=of q_0] {$q_1$};
  \path[->] (q_0) edge                node [above] {a(+), s(+), v(+)} (q_1)
                  edge [loop above]   node         {a(-), s(-), v(-)} ()
            (q_1) edge [bend left]    node [above] {s(-)} (q_0)
                  edge [loop above]   node         {a(+), s(+), v(+), a(-), v(-)} ();
\end{tikzpicture}
\caption{DFA encoding of the first policy of Section~\ref{sec:overview-example}.}
\label{fig:dfa}
\end{figure}

The resulting automaton is depicted in Figure~\ref{fig:dfa}.
We use an incoming arrow and a double circle to denote initial and final states, respectively.
\end{example}

A trace $\sigma$ \emph{complies} with a policy if and only if it is accepted by the automaton generated from the policy.
For instance, the trace $\sigma = v(+)s(-)a(-)$ is accepted by the automaton of Example~\ref{ex:semantics}, while $\sigma' = v(+)s(-)a(+)$ is rejected.

\section{Privacy preserving enforcement}\label{sec:details}\acnote{ma il titolo e' giusto?}

%


\paragraph{Oblivious transfer.}

For the policy evaluation we adopt the protocol presented in~\cite{ob-automata}.
The protocol is executed between two parties, privately holding a DFA $A$ and an input trace $\sigma$, respectively.
The goal of the protocol is to allow the two parties to check whether $A$ accepts $\sigma$, while keeping them private.

The overall protocol consists of three sub-protocols.
The first protocol handles the first transition of the automaton from the initial state and configures the protocol parameters for the next step.
The second one implements a single step of the transition function by means of homomorphic encryption in order to hide both the last taken transition and the current state of $A$ (which could leak information about the last element of $\sigma$).
When $\sigma$ reduces to the empty trace, the last protocol checks if the current state belongs to the set of final ones.

It is worth noticing that the complexity of the protocol above is linear in $(i)$ the size of $\Lambda$, $(ii)$ the size of $Q$ and $(iii)$ the length of $\sigma$. 
$\Lambda$ and $Q$ are known in advance by the policy owner, who can estimate the policy evaluation time.
On the other hand, excessive dimensions of $\sigma$ can be avoided, e.g., by truncating the trails of older events that are interesting no longer.
For instance, this can be implemented by means of an authoritative participant playing the role of a certified clock that triggers a day-change event every 24 hours.

\paragraph{Token generation and verification.}

Once a trace is accepted by the third sub-protocol above, the token can be generated. 
A simple way to generate it consists in taking the hash of the accepted trace and sign it using the private key of the policy owner, a.k.a. the facility. 
To prevent participants from replaying a token, an expiration date can be added to the signature.
The validity of an authorization token can be checked by any party knowing the public key of the token issuer.

%

\paragraph{Contact oracle server.}

The main difference between policies 1 and 2 of Section~\ref{sec:overview-example} is that the former only evaluates events referring to a single user, i.e., the outcome of its clinical tests, while the latter concerns also dangerous contacts.
By definition, a contact is an event involving two users, e.g., $u$ and $w$.
Thus, a policy evaluation of the trace of $u$ might indirectly leak $w$'s private data.

To avoid this, we introduce the \emph{contact oracle server}.
The oracle stores a list of anonymous identifiers corresponding to users that were discovered positive to COVID-19.
Such a list is continuously updated by the authoritative participants.

Policy 2 can be implemented by considering a special event $d$ denoting a dangerous contact.
However, such an event cannot appear in any user's trace, since users cannot distinguish this kind of contacts.
Instead, traces contain generic contact events of the form $c(i)$, where $i$ is the anonymous identifier of the user we had a contact with, e.g., $i$ entered the Bluetooth transmission range.

Before a policy evaluation starts, the policy owner queries the oracle for the list of dangerous contacts identifier.
In every policy rule that applies to $d$, $d$ is replaced with the list of events $c(i)$, for each $i$ appearing in the oracle database.

\section{Related work}
\label{sec:literature}

To the best of our knowledge we are the first to propose a history-based policy framework for contact tracing.
Still, the working assumptions of our proposal are similar to those of other system for contract tracing.
Here, we describe the proposals more similar to ours and we refer the interested reader to~\cite{tang2020privacy}.

In~\cite{troncoso2020decentralized,reichert2020privacy} two fully private contact tracing systems are proposed.
These proposals aim only at tracking the contact of a user and to detect possible dangerous encounters achieve full privacy.
So, they do  not provide any means through which a public authority can check that a person is compliant with the request to stay in quarantine.
Our proposal instead is an authorization infrastructure that allows defining policies and checks their compliance in a privacy preserving manner.

Brack et al.~\cite{brack2020decentralized} propose a \emph{distributed hash table}  where every user can modify her own entries and submit anonymous queries to check the entries of her contacts.
Differently from our work, their proposal focuses on sharing information among the users for favoring responsible behavior and provides no support for the definition of fine-grained policies.

Micali~\cite{Micali} proposes that
the contact data collected by a cellphone are redacted (for privacy reasons), then simultaneously uploaded on a server belonging to the health authority, hashed and stored on a blockchain.
In this way, the authority can track the evolution of the epidemic and decide the most appropriate response, and the citizens can verify the authenticity of the used data.
Differently from our proposal, there is no mechanism for  the authority to checks that citizens are compliant to its regulations.

\section{Concluding remarks}
\label{sec:issues}

The collection of personal data respects privacy whenever either there is an explicit consent, or there are ``some other legitimate basis laid down by law.''
\footnote{See article 5.2 of the ``Convention 108+'', a text developed by the Council of Europe currently signed by 55 countries, including several non European. See \url{https://www.coe.int/en/web/conventions/full-list/-/conventions/treaty/108}. }   
In the second case, the collection of personal data should be ``adequate, relevant and not excessive in relation to the purposes for which they are processed''.\footnote{Ibid. Article 5.c.}
Hence, when personal data are automatically collected  privacy is achieved by limiting it to people or situations deemed to pose a risk.
A typical example is speed cameras, where personal data (i.e., license plate, location, time of the day, speed) is recorded \emph{only} for those driving above the speed limit --- those posing a risk to themselves and to others. 
Combining privacy with enforceability is even more challenging. 
In \projname{}, the personal data stored in each trace is never disclosed. 
The only piece of information that can be revealed to a public authority is the authorization to access a certain facility. 
However, under reasonable conditions, knowing whether a user is authorized to enter a certain facility is a critical information for the public safety. 

Open issues that we plan to investigate as future work are $(i)$ the efficiency of practical implementation, $(ii)$ the relevant attacker model, $(iii)$ the extension to global policies, and $(iv)$ the integration of an incentives system.

Although the proposed solution is theoretically efficient (see Section~\ref{sec:details}), our oracle-based approach significantly enlarges the size of the policy alphabet.
Experiments are needed to assess this possible limitation.

The attacker model is also crucial ti establish the formal properties of our proposal.
In particular, we plan to consider dishonest protocol participants, e.g., free riders and curious users.

Our CPSL is fairly expressive, but cannot model policies that predicate over the behavior of more than one user.
In certain cases, we might want to apply policies that depend on the context, e.g., at most 1 user that had dangerous contacts can access.
This kind of policies require to globally evaluate the trace of many, possibly all the participants. 

Finally, we believe that the success of contact tracing systems strongly relies on appropriate subsidies. 
For example, once a token is revoked, a user can receive a transfer for compensating her for the quarantine costs.
These monetary transfers can also depend on the aggregate state of the system, for example on the overall number of users under quarantine.
Integration with existing technologies such as cryptocurrencies can automate the transfer mechanism.

\bibliographystyle{splncs04}
\bibliography{biblio}

\end{document}